# Factors Influencing Autonomously Generated 3D Geophysical Spatial Models

M. Balamurali[1*], A. Melkumyan[1] and J. Zigman[1]

[1] Australian Centre for Field Robotics, University of Sydney, Australia
mehala.balamurali@sydney.edu.au

Understanding the contribution of geophysical variables is vital for identifying the ore indicator regions. Both magnetometry and gamma-rays are used to identify the geophysical signatures of the rocks. Density is another key variable for tonnage estimation in mining and needs to be re-estimated in areas of change when a boundary update has been conducted. Modelling these geophysical variables in 3D will enable investigate the properties of the rocks and improve our understanding of the ore. Gaussian Process (GP) was previously used to generate 3D spatial models for grade estimation using geochemical assays. This study investigates the influence of the following two factors on the GP-based autonomously generated 3D geophysical models: the resolution of the input data and the number of nearest samples used in the training process. A case study was conducted on a typical Hammersley Ranges iron ore deposit using geophysical logs, including density, collected from the exploration holes.

## INTRODUCTION

Autonomous Gaussian Process (GP) was used to develop 3D spatial models for grade estimation using information obtained from exploration drilling and supported by other field data such as surface mapping (Melkumyan, A., Ramos, F.T., 2009). Exploration holes are often drilled through to the base of the deposit (up to 200 m) and spaced at intervals sufficient to allow the lateral variations in the geology to be revealed. As the data is horizontally sparse (~50m) and vertically dense (10cm between the samples) and different data densities can be present in different parts of the mine, it is important to develop an autonomous robust approach for choosing the data for estimation of each block in the model in order to produce accurate probabilistic grade models. Previous studies were conducted by the authors to identify the mineralised intervals in the drill holes using GP on geochemical assays (Balamurali, M. and Melkumyan, A. 2014, 2015).

However, application of GP to geochemical assay data (representing 1.5-2m interval of material) is different to application of GP to geophysical measurements. Probabilistic volumetric estimates for geophysical data can be generated using composited (e.g. 2m interval) as well as high-resolution (10cm interval) data. Thus, it is vital to identify the best compositing length that will allow better GP based 3D modelling so that the information conveyed by the models and the production risks are improved.

In addition to the resolution of the input data, another factor impacting the 3D spatial model is the number of samples used for inferencing. The local neighbourhood search is aimed at using local nearest data to inform the inference of a block. This procedure is similar to the local neighbourhood search in Kriging (Vann et al,2003, Rivoirard, 1987). The observed values at exploration holes are used to estimate the grade in unmeasured locations using GP. Search radius plays a critical role in choosing the best neighbours of a given block as it defines the maximum spatial volume that contributes to the inference of a block. The minimum number of samples that are needed for inferencing the block value is another important parameter to be analysed (Neufeld and Wilde, 2005).

KD-tree or octant nearest neighbour search methods are widely used in  conventional grade control models. In the octant based search, the search ellipse is divided into eight equal-angle sectors while in

KD-tree the search spaces is divided in to two. The purpose of this study is to ensure that each block value is inferred from the best possible values of its neighbours and develop an understanding of the key factors influencing the GP-based autonomously generated models.

This study analyses the learning and inferencing process of an autonomous GP using different compositing lengths: from 2m length composite down to 0.1m length natural gamma, magnetic susceptibility and density data. A case study was conducted on a typical Hammersley Ranges iron ore deposit. Geophysical logs were collected for the exploration holes. GP was used to learn the hyperparameters (length scales in x, y and z directions, amplitude and noise) for geophisical exploration data in the test region using several different composting lengths. The properties of the resulting models for each of the compositing lengths were evaluated through analysis of the mean and variance estimated for each block in the block model. The proposed approach enables to identify the best compositing length for a given modelling task.

This study also evaluates the exploration-only density modelling based domain interpretation quality, by incorporating KD-tree or octant neighbourhood search methods with the autonomous Gaussian Process (GP). Comparison between different neighbourhood search methods will allow better 3D model for the 3D spatial distribution of the data. Histograms and spatial distribution plots of the data were used to identify and address the causes of having output models with different characteristics.

**DATA AND METHODOLOGY**

**Data**

Experiments were performed based on information obtained from drilling at the test deposit of a typical Marra Mamba style banded iron formation (BIF) hosted deposit in the Hamersley Region of Western Australia. Geophysical logs, including natural gamma, Mag-sus and density, were collected for the exploration holes. The holes are generally 25-100m apart and tens to hundreds of meters deep. Within each hole, data is collected at 0.1m interval and then composited into 0.2m, 0.3m, 0.4m, 0.5m, 0.6m, 1m and 2m intervals. The measurements include the position (east, north, elevation) data along with the corresponding geological domain and the drill-hole name.

**Gaussian Process**

GPs provide a probabilistic method of modelling functions representing quantities of interest (e.g. weight percentage of chemical elements, mineralogy, etc.) given a set of data. Mathematically a GP is an infinite collection of random variables, any finite number of which has a joint Gaussian distribution. Machine learning using GPs consists of two steps: training and inference. GPs usually contain unknown hyper-parameters and the training step is aimed at optimising those hyper-parameters to result in a probabilistic model that best represents the training library. The hyper-parameters used in the applications of GPs include the length scales, which describe the rate of change, and the noise variance, which describes the amount of noise in the data set. Once the hyper parameters are known they can be used during the inference step to predict the values of the function of interest at new locations.

Consider a training set $D = (X, y)$ consisting of a matrix of training data $X = [x_1; x_2; \ldots; x_N]^T$, where $N$ is the number of samples, $T$ indicates a transposed vector or matrix and a corresponding target vector $y = [y_1; y_2; \ldots; y_N]^T$. To each vector $X_i \in \mathbb{R}^D, (i = 1, 2, \ldots, N)$, we associate a target value $y_i \in \{-1, 1\}$. The aim is to compute the predictive distribution $f(x_*)$ at a new test point $x_*$. A GP model places a multivariate Gaussian distribution over the space of function variables $f(x)$, mapping input to output spaces. GPs can also be considered as a stochastic process which can be fully specified by its mean function $m(x)$ and its covariance function $k(x, x')$. To completely describe the standard regression model we assume Gaussian noise $\varepsilon$ with variance $\sigma^2$, so that $y = f(x) + \varepsilon$. With the training set $(X, f, y) = (\{x_i\}, \{f_i\}, \{y_i\})_{i=1:N}$ and test set $(X_*, f_*, y_*) = (\{x_{*i}\}, \{f_{*i}\}, \{y_{*i}\})_{i=1:N}$ and with $m(x) = 0$, the joint distribution becomes:

$$\begin{bmatrix} y \\ f \end{bmatrix} \sim N\left(0, \begin{matrix} K(X,X) + \sigma^2 I & K(X,X_*) \\ K(X,X_*) & K(X_*,X_*) \end{matrix}\right) \tag{1}$$

In Equation 1, $N(\mu, cov(f_*))$ is a multivariate Gaussian distribution with mean µ and covariance $cov(f_*)$ and K is the covariance matrix computed between all the points in the set. By conditioning on the observed training points the predictive distribution for new points can be obtained as:

$$p(f_*|X_*, X, y) = N(\mu, cov(f_*))$$

where:

$$\mu = K(X_*, X)[K(X, X) + \sigma^2 I]^{-1} y, \tag{2}$$

$$cov(f_*) = K(X_*, X_*) - K(X_*, X)[K(X, X) + \sigma^2 I]^{-1} K(X, X_*) \tag{3}$$

Learning a GP model is equivalent to learning the hyper parameters of the covariance function from a data set. In a Bayesian framework this can be performed by maximising the log of the marginal likelihood with respect to $\theta$:

$$\log p(y|X, \theta) = -\frac{1}{2} y^T [K(X,X) + \sigma^2 I]^{-1} y - \frac{1}{2} \log[K(X,X) + \sigma^2 I] - \frac{N}{2} \log 2\pi \tag{4}$$

The marginal likelihood has three terms (from left to right) that represent the data fit, complexity penalty (to include the Occam's razor principle) and a normalisation constant. It is a non-convex function on the hyper parameters $\theta$ and therefore only local maxima can be obtained. Good local maxima can be obtained using gradient descent techniques by using multiple starting points.

For the tasks of this paper multiple length scale squared exponential covariance function was used

$$k(x, x', \Sigma) = \sigma_f^2 \cdot exp\left[-\frac{1}{2}(x - x')^T \Sigma (x - x')\right] \tag{5}$$

For example, in the context of geological grade estimation using ten chemical variables, $\Sigma$ can be chosen to be a d × d diagonal length scale matrix (d is the dimensionality of the inputs, which is equal to ten in this case); $\sigma_f^2$ is the signal variance.

More information on GPs can be found in Rasmussen and Williams (2006) and Williams C. K. I. and Rasmussen C. E. (1996).

**Nearest Neighbor Search Method**

Octant and KD-tree based neighbourhood search methods were applied to the exploration-only density data and the results were compared. If octant search is enabled, the search space is divided into eight equal-angle sectors, and only the specified maximum number of points (max-per-octant) from each octant will be used. There is no set rule for minimum number of octant to be chosen. KD-tree method allows both a minimum number of training points, and a radius size to be set. If less than the minimum number of training points is returned then the algorithm perform a nearest neighbour search for the minimum number of training points required.

Histograms and spatial distribution plots of the data and model were produced for each models and the differences were analysed to identify and address the causes of having output models with different characteristics.

The basic inference steps on a given block are as below:
1. Choose the block to find the grade estimation.
2. Find the nearest exploration data to that block using nearest neighbourhood search algorithm.

3. Calculate the covariance of the nearest samples.
4. Calculate the grade of the block.

Similarly all blocks' values were calculated.

## RESULTS AND DISCUSSION

### Impact of geophysical data compositing

GP was used to learn the hyper-parameters (length scales in x, y and z direction, amplitude and noise) for gamma, Mag-sus and density values using the exploration data in the test region at different composting lengths. As it can be seen in Figure 1, the length scale of z continuously decreases with the increasing resolution of training data. Also, significant changes can be observed in the hyper-parameters when the compositing length becomes smaller than 0.6m.

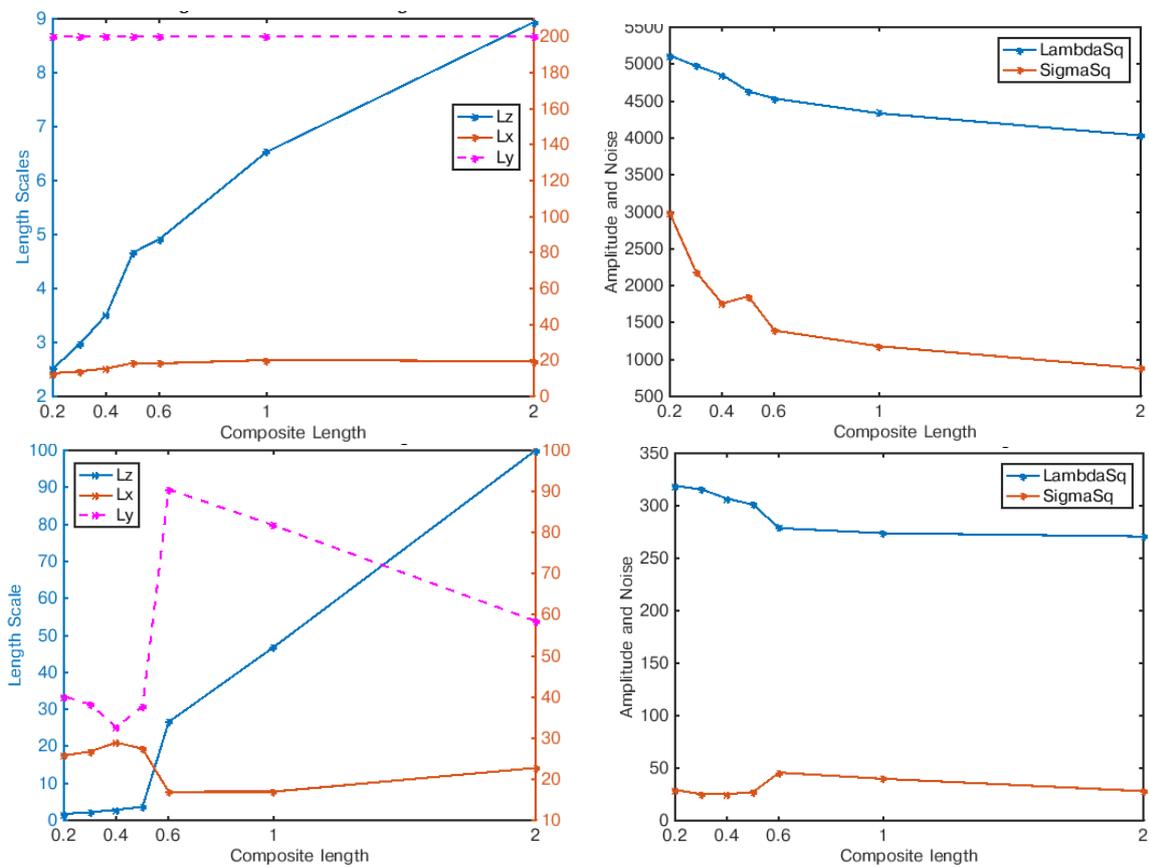

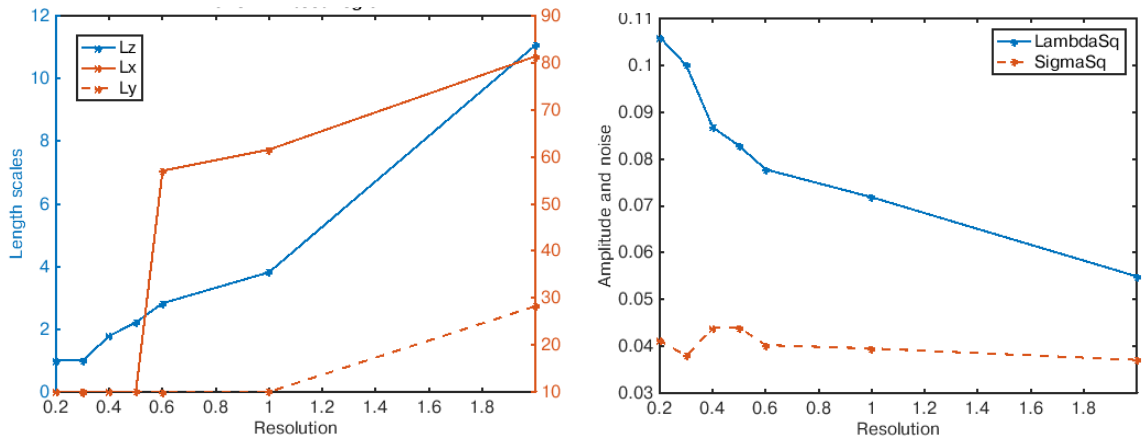

*Figure 1 column 1: The length scales; column 2: the amplitude and noise at different composite lengths learned by GP using Mag-Sus (raw 1), Gamma (raw 2) and density (raw 3)*

Then the autonomously learned hyper-parameters were used to predict the Mag-sus, gamma and density values for blocks with dimension of 20*20*5 using GP. Figure 2 shows the predicted mean and variance of each estimated block for different composited lengths applied to the Mag-Sus and Gamma data (training samples). It can be observed from Figure 2a that the behaviour of the model changes once the compositing length becomes smaller than 0.6m. The result shows that the high-resolution data impacted the learned parameters (length scales and noise parameters) and the overall inference when training samples of different compositing lengths were used. Figure 2b shows similar impact on the estimated variance. 0.6m in again the compositing length below which the significant jump happens, in this case collapsing the inference variance.

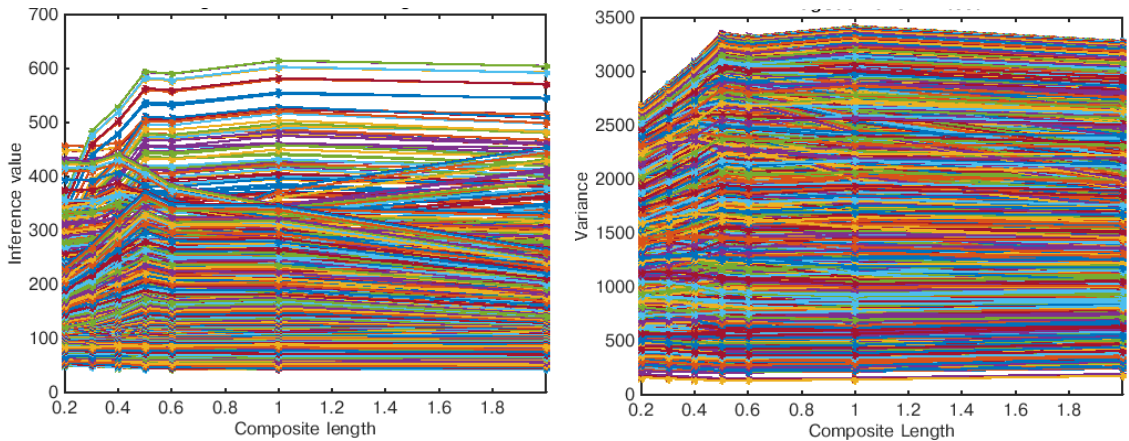

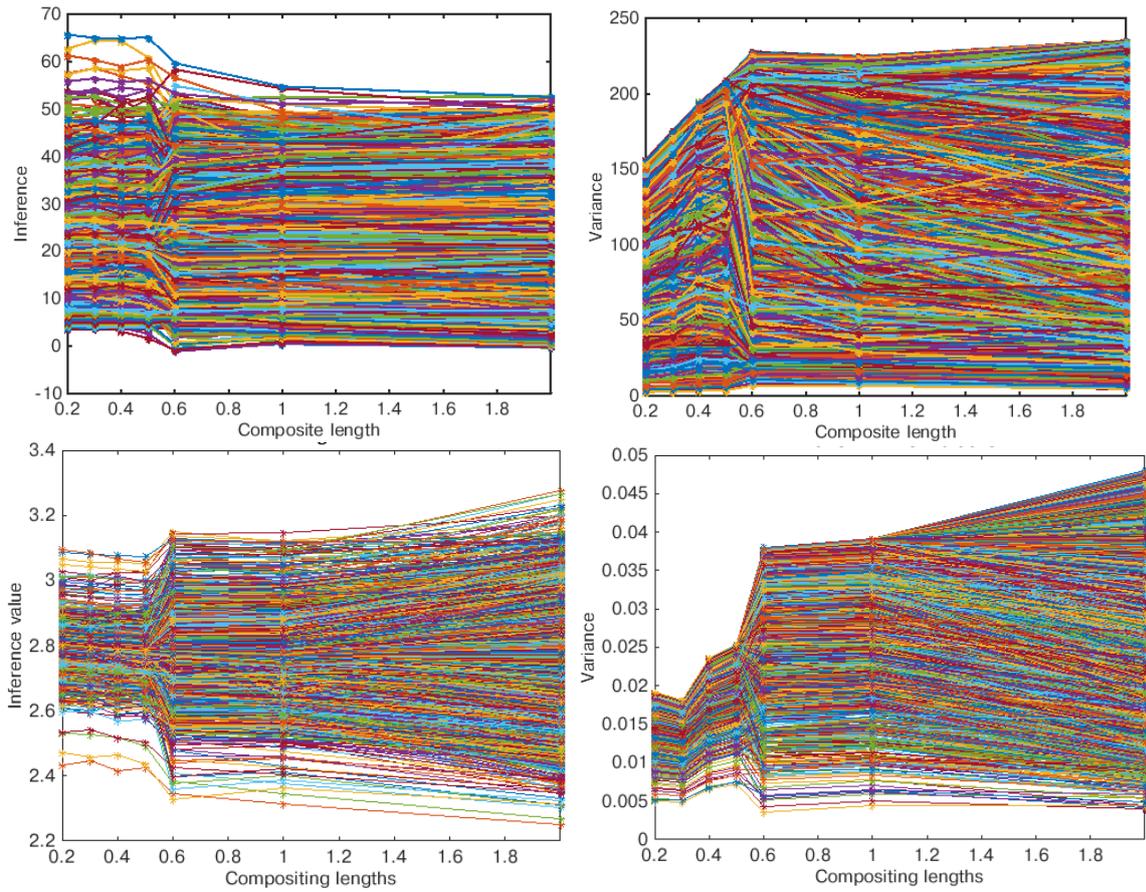

*Figure 2: (a) inference and (b) variance of each block predicted by GP using Mag-Sus (raw 1), Gamma (raw 2) and density (raw 3).*

Figure 3 shows the slices of 3D spatial model of density at different compositing length and there are significant difference between the density distributions in the 3D geological domain. That is the density information is not well propagated through as we get below 0.6. This is concerning because it has been ensured that the input data used for the models are accurately find the correct density information 3D spatial models. Analysing these differences has revealed that their significant portion is due to the difference in the resolution of input data for the blocks that are not well informed.

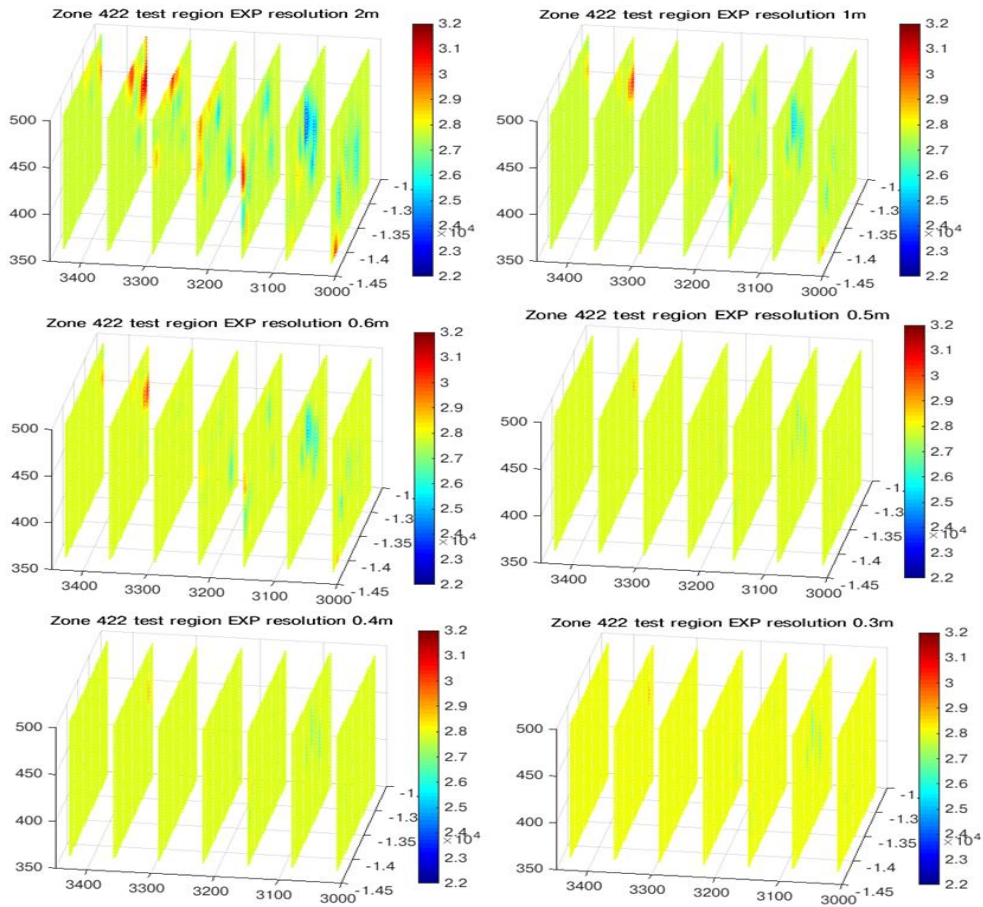

*Figure 3: Cross-sections through the model and the figure shows that the density information is not well propagated through as we get below 0.6.*

Results demonstrate that the length scales for all the spatial directions autonomously learned by the GP decrease as the compositing length decreases (i.e. as the resolution of the data increases). The results also indicate the presence of a threshold for the compositing length beyond which the length scales become too small to propagate information between the drill holes impacting the quality of the resulting 3D model.

**Impact of the neighbourhood selection strategy**

Another potential factor that can influence the inference results is number of nearest neighbours used in training process. The above results were obatined by using all measured exploration data in training purpose. However, the number of training samples were decreased with increasing composited length scales. In order to usderstand the impact on inferencing on block using different number of nearest neighbours used by each block, another study was coducted on a different test region using denisty only data with constant high-resolution data.

As it can be observed in the 3D spatial plots in Figure 4, there are significant difference between the density distributions in the 3D geological domain. The artefact by these models is clearly shown by the red oval shape. This is concerning because it has been ensured that the input data used for both models are the same. Analysing these differences has revealed that their significant portion is due to the difference in the modelling approach for the blocks that are not well informed.

Nearest neighbour samples were chosen using two different methods. First method used octant search and during each search iteration number of samples obtained from each octant is controlled by maximum samples per octant. As can be seen in the histograms in Figure 4, the difference between the artefacts in both 3D plots of mineralised Domain A are due to the different number of nearest neighbours used by each block. Left histogram (case 1) of Figure 4 shows, when the search through octant is enabled, the search ellipse is divided into eight equal-angle sectors, and only the specified maximum number of points from each octant were used. This limits the number of samples used from each octant and hence limits the total maximum. However, there is no set rule for minimum number of octant to be chosen. Thus, the model allows any number of nearest neighbour samples between zeros to maximum for each block. Here the maximum number of samples are automatically limited by the max-per-octant parameter.

In contrast to this, the second method allows both a minimum number of training points, and a radius size to be set. In this case a radius search is performed first, and if it yields enough points, then those for the training set. However, if less than the minimum number of training points is returned then the model perform a nearest neighbour search for the minimum number of training points required. It can be clearly seen in the histogram at right side of Figure 4. As there were not enough nearest neighbour samples in the given ellipsoid, the second method picked the minimum number of samples for each block. In this case the the minimum number of samples is 30. 3D spatial plots in last row of Figure 4 show an example of nearest data for a particular block.

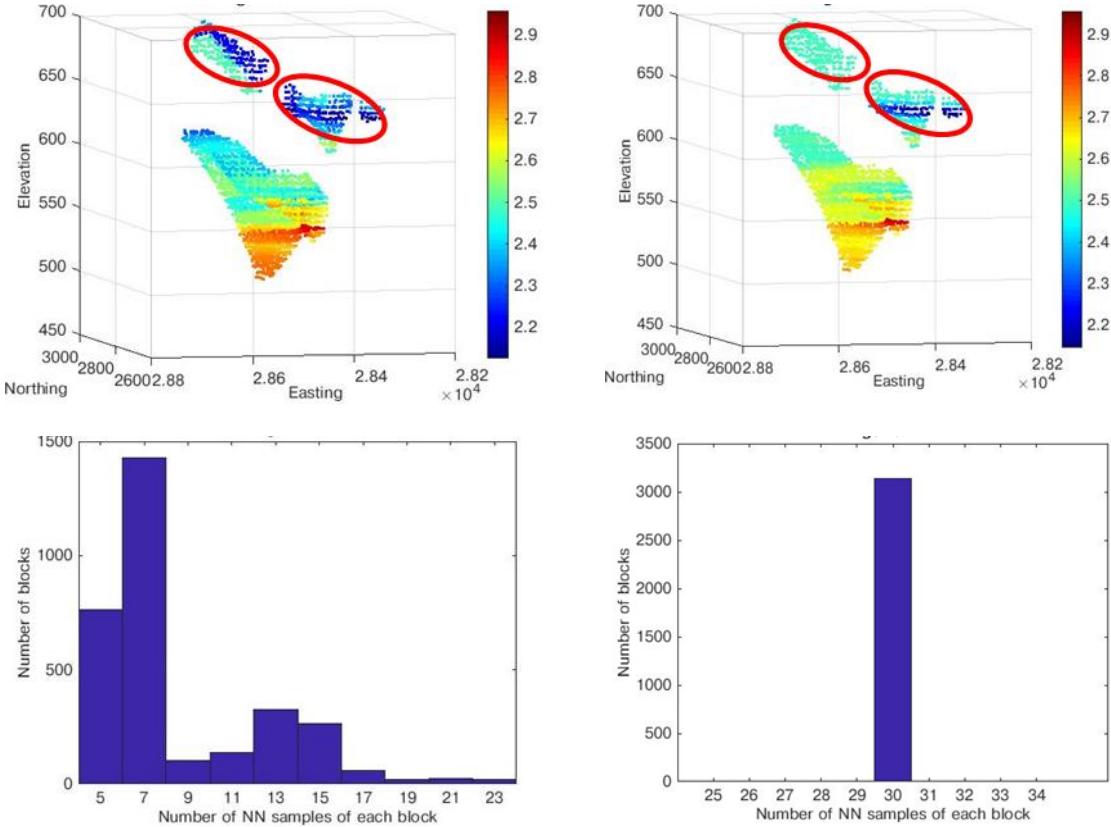

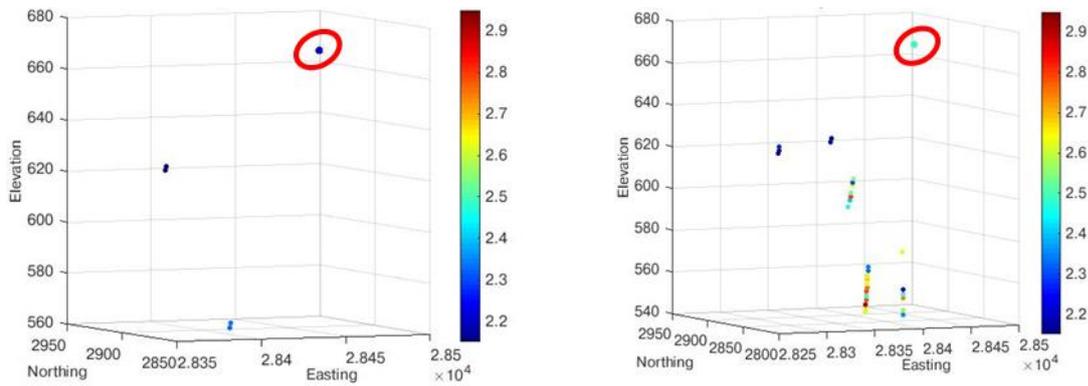

*Figure 4. A small amount of high-density values in two single hole in the Domain A has impacted a large area in the model (left column) by method 1. In the 3D spatial plot on right side, a small amount of high-density values in two holes in the same Domain A has limited area of influence in the model 2 as it is using other NN data from different holes. The number of nearest neighbours used by each block can be seen in the corresponding histograms. The 3D spatial plots in the last row show an example block with inferred value (circled with red) and their corresponding nearest neighbours for both model 1 and 2.*

Another study was conducted on another mineralised test region (Domain B). Compared to previous test region, this region has abandoned number of exploration-hole density data. As can be seen in Figure 5, the inferenced valuess by the models using both nearest neighbour search methods are significantly close. This is because of the number of NN samples obtained by each models are adequate for accurate prediction. This can be explained by the histograms of number of NN samples used by each test samples.

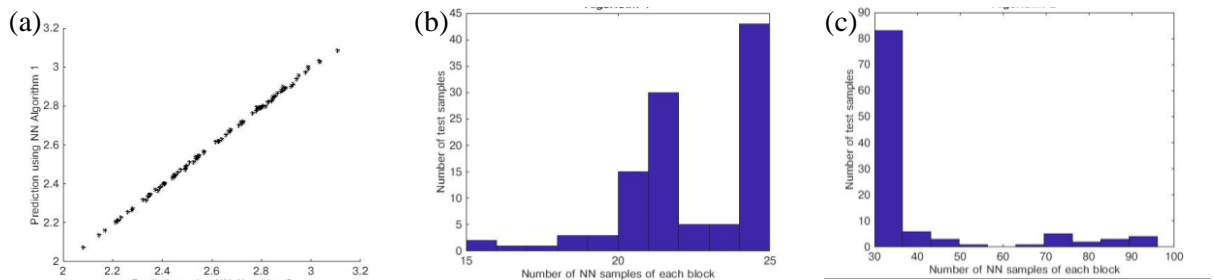

*Figure 5. Scatter plot (a) show the agreement of prediction by the models using the Octant and KD-Tree. The number of nearest neighbours used by each block can be seen in the corresponding histograms from the model 1(b) and model 2 (c)*

This study revels one of the major factor influencing the end results by the models using different search methods. This can be often present in real situations when there is not enough samples to model. This affects can be mainly seen at boundaries of domain. Experiment on Domain B was used as a validation model and the results show that this affect is very low when there is enough training samples. However, further analysis is required to reliably develop 3D spatial models for grade estimation.

# CONCLUSION

This study analysed the learning and inferencing outcomes of an autonomous Gaussian Process (GP) for developing 3D spatial models of geophiysical data using different compositing lengths and artiffects created by different NN serach strategy. The outcomes of the autonomous GP show that the length scales learned in each direction decrease as the compositing length decreases (i.e. as the resolution of the data increases). The results also show a transition point in the compositing length below which the learned length scales sharply decrease.

The dynamics of the resulting models for each of the compositing lengths were evaluated through the analysis of the mean and variance estimated for each block in the block model. The results show that information is not well propagated between exploration holes when the compositing length goes below the transition point. The proposed approach enables to identify the best compositing length for a given modelling task.

Furthermore, the comparison presented in this study shows that the block inference values by the GP can be significantly impacted by the nearest neighbour samples chosen by different NN search methods. However, this artefact is mainly happened at model boundaries or when there is not enough exploration hole data. This is evident by the results obtained in domain with abondoned data samples where there is enough NN samples can be selected by the methods presented here. The results particularly demonstrate that different neighbourhood selection strategies can lead to different artefacts that impact the quality of the models, especially in areas that are poorly informed by the available data.

This study revels the major factor influences the autonomous 3D spatial models using geophysical data that can be often present in real situations. However, more analysis is required to reliably develop 3D spatial models.

# ACKNOWLEDGEMENT

This work has been supported by the Australian Centre for Field Robotics and the Rio Tinto Centre for Mine Automation.